\catcode`\@=11

\@addtoreset{equation}{section}

\def\@normalsize{\@setsize\normalsize{15pt}\xiipt\@xiipt
\abovedisplayskip 14pt plus3pt minus3pt%
\belowdisplayskip \abovedisplayskip
\abovedisplayshortskip  \z@ plus3pt%
\belowdisplayshortskip  7pt plus3.5pt minus0pt}

\def\small{\@setsize\small{13.6pt}\xipt\@xipt
\abovedisplayskip 13pt plus3pt minus3pt%
\belowdisplayskip \abovedisplayskip
\abovedisplayshortskip  \z@ plus3pt%
\belowdisplayshortskip  7pt plus3.5pt minus0pt

\def\@listi{\parsep 4.5pt plus 2pt minus 1pt
            \itemsep \parsep
            \topsep 9pt plus 3pt minus 3pt}}

\def\underline#1{\relax\ifmmode\@@underline#1\else
        $\@@underline{\hbox{#1}}$\relax\fi}
\@twosidetrue

\catcode`@=12

\catcode`\@=11

\catcode`\@=12

\relax

\def\figcap{\section*{Figure Captions\markboth
        {FIGURECAPTIONS}{FIGURECAPTIONS}}\list
        {Fig. \arabic{enumi}:\hfill}{\settowidth\labelwidth{Fig. 999:}
        \leftmargin\labelwidth
        \advance\leftmargin\labelsep\usecounter{enumi}}}
 \relax
\def\tablecap{\section*{Table Captions\markboth
        {TABLECAPTIONS}{TABLECAPTIONS}}\list
        {Table \arabic{enumi}:\hfill}{\settowidth\labelwidth{Table 999:}
        \leftmargin\labelwidth
        \advance\leftmargin\labelsep\usecounter{enumi}}}
 \relax
\def\reflist{\subsubsection*{References\markboth
        {REFLIST}{REFLIST}}\list
        {[\arabic{enumi}]\hfill}{\settowidth\labelwidth{[999]}
        \leftmargin\labelwidth
        \advance\leftmargin\labelsep\usecounter{enumi}}}
 \relax

\catcode`\@=11

\def\FERMIPUB{}

\def\ps@headings{\def\@oddfoot{}\def\@evenfoot{}
\def\@oddhead{\hbox{}\hfill
        \makebox[.5\textwidth]{\raggedright\ignorespaces --\thepage{}--
        \hfill {\rm FERMILAB--Pub--\FERMIPUB}}}
\def\@evenhead{\@oddhead}
\def\subsectionmark##1{\markboth{##1}{}}
}

\ps@headings

\relax
\newskip\humongous \humongous=0pt plus 1000pt minus 1000pt

\newif\ifdtup

\def\beq{\begin{equation}}
\def\eeq{\end{equation}}

\def\beqn{\begin{eqnarray}}
\def\eeqn{\end{eqnarray}}
\relax

\def\dotx{\dotx{\dot\overline{x}}}

\documentstyle[art12,A4]{article}
\begin{document}
\begin{titlepage}
\begin{flushright}
Z\"urich University Preprint\\
ZU-TH 22/94\\
\end{flushright}
\begin{center}
\vfill
{\large\bf ASYMPTOTIC BEHAVIOR OF COMPLEX SCALAR
FIELDS IN A FRIEDMAN-LEMAITRE UNIVERSE$^*$}
\vfill
{\bf David Scialom and Philippe Jetzer}
\vskip 0.5cm
Institute of Theoretical Physics, University of Z\"urich,
Winterthurerstrasse 190,\\CH-8057 Z\"urich, Switzerland
\end{center}
\vfill
\begin{center}
{\bf Abstract}
\end{center}
\begin{quote}
We study the coupled Einstein-Klein-Gordon equations for
a complex scalar field with and 
without a quartic self-interaction in a curvatureless
Friedman-Lema\^{\i}\-tre Universe.  
The equations can be written as a set of 
four coupled 
first order non-linear differential equations, for which 
we establish the phase portrait for the time evolution of the scalar
field.
To that purpose we find the singular points of the differential 
equations lying in the finite region and at infinity of the phase space
and study the corresponding asymptotic behavior of the solutions. 
This knowledge 
is of relevance, since it provides the initial conditions which are needed
to solve numerically the differential equations.
For some singular points lying
at infinity we recover the expected emergence of an 
inflationary stage. 

\end{quote}
\vfill
\begin{center}
October 1994 
\end{center}
\vfill
$^*$ This work was supported by the Swiss National Science
Foundation.\\
 
\end{titlepage}
\newpage
\noindent{\bf 1. INTRODUCTION}\\

The recent developments in particle physics and cosmology
suggest that scalar fields may have played an important role in the 
evolution of the early Universe, for instance in primordial
phase transitions, and that they may constitute part of the dark matter.
Moreover, scalar fields are predicted by most of the particle physics models
based on the unification of the fundamental forces, as for instance
in superstring
theories. Scalar particles are needed in cosmological
models based on inflation, whose relevance is supported by the
results of the COBE-DMR measurements that are consistent with
an Harrison-Zel'dovich (scale invariant n=1) spectrum \cite{kn:Smoot}.
These facts, in particular inflation, 
motivated the study of the coupled Einstein-scalar field
equations to determine the time evolution and also
the gravitational equilibrium configurations of the scalar fields.
The latter one in particular for massive complex scalar fields,
which may form so-called boson stars \cite{kn:Jetzer,kn:Liddle}.
  
A detailed study of the solutions of the Einstein equations
for a homogeneous isotropic Friedman-Lema\^{\i}tre Universe with
a real scalar field has been done in particular
by Belinsky et al. \cite{bel1,bel2,bel3}, see also ref.\cite{kn:Piran}.
In this paper we extend, following ref.\cite{bel1,bel2,bel3}, 
these investigations to a complex scalar field.  

This analysis is important in order to see the degree of generality of
solutions possessing an inflationary stage and also due to the fact that
these solutions constitute the background, starting from which
one can study in the early Universe
the time evolution of perturbations for the 
scalar field \cite{kn:Deruelle,kn:Mukhanov}. 
A fact this, which is of relevance if scalar fields
make up part of the dark matter. If this is the case they may also form
compact objects, such as bose stars, or trigger the formation
of observed large scale structures in the Universe. 
 
Since we consider a complex scalar field with or
without a quartic self-interaction in a curvatureless
Friedman-Lema\^{\i}tre Universe, we get for the Einstein-Klein Gordon
equations a set of four  
first order non-linear differential equations, for which
we study the phase portrait for the time evolution of the
scalar field. We first determine the singular 
points of the differential equations 
and then find analytically the asymptotic behavior for the solutions
nearby these points. 
For the singular point lying in a finite region of the phase space, 
we can use for $m \neq 0$, in an adapted coordinate system, the averaging
method in order to get the asymptotic behavior
of the solution, whereas, for the points lying at infinity, we
first compactify the phase space on the lower hemisphere of the
3-dimensional sphere. We can then 
apply the Poincar\'{e}-Dulac theorem to get the corresponding
asymptotic behavior. From the solution 
we then see if there is inflation and how long it
lasts. For some singular points lying
at infinity, we recover the expected emergence of an 
inflationary stage. 

The paper is organized as follows: in section 2 we present the basic
equations which we will use. In section 3 we first study for a massive scalar
field the singular point
lying in the finite region of the phase space and then the singular
points lying at infinity both with and without a quartic self-interaction.
Section 4 is devoted to the massless scalar field with a quartic
self-interaction and a short summary concludes the paper.\\

\noindent{\bf 2. BASIC EQUATIONS}\\

We consider a massive complex scalar field with quartic self-interaction
in a Friedman-Lema\^{\i}tre Universe with the following action 
\begin{equation}
S=\int (-\frac{R}{16\pi G}+ e_{\mu}\varphi e^{\mu}\varphi^{\ast}
+m^{2}\varphi \varphi^{\ast}+\lambda
(\varphi \varphi^{\ast})^{2})\sqrt{-g}~d^{4}x~,
\end{equation}
where $g$ is the determinant of the metric
\begin{equation}
ds^2=g_{\mu\nu}\theta^{\mu}\theta^{\nu}=-\theta^{o}\theta^{o}+
\delta_{ij}\theta^{i}\theta^{j}~,
\end{equation}
with
\begin{displaymath}
\theta^{o}=dt\mbox{ , } \theta^{i}=
\frac{a(t) dx^i}{(1+\frac{k}{4}r^{2})}~,\mbox{ }r^{2}=\sum_{i=1}^{3} x_{i}^{2}
\end{displaymath}
and $e_{\mu}$ is the dual basis of $\theta^{\mu}$.
By varying the action with respect to $g^{\mu\nu}$ we get the
Einstein field equation
\begin{equation}
G_{\mu\nu}=8\pi G T_{\mu\nu}~,   \label{ein}
\end{equation}
with
\begin{equation}
T_{\mu\nu}=e_{\mu}\varphi e_{\nu}\varphi^{\ast}+e_{\nu}\varphi
e_{\mu}\varphi^{\ast}-g_{\mu\nu}(g^{\alpha\beta}e_{\alpha}
\varphi e_{\beta}\varphi^{\ast}+m^2\varphi
\varphi^{\ast}+\lambda (\varphi \varphi^{\ast})^{2})~. 
\end{equation}
The (00) component of eq.(\ref{ein}) leads to the
constraint equation
\begin{equation}
H^{2}+\frac{k}{a^{2}}=\frac{8\pi G}{3}(\dot{\varphi}
\dot{\varphi}^{\ast}+m^{2}\varphi \varphi^{\ast}+\lambda
(\varphi \varphi^{\ast})^{2})~,   \label{e00}
\end{equation}
with $H=\dot{a}/a$ and dot means derivative with respect to time.
For the (ij) component we have
\begin{equation}
(-2\dot{H}-3H^{2}-\frac{k}{a^{2}})\delta_{ij}=8\pi G
( \dot{\varphi}\dot{\varphi}^{\ast}-m^{2}\varphi \varphi^{\ast}-
\lambda (\varphi \varphi^{\ast})^{2})\delta_{ij}~.
\label{eij}
\end{equation}
By varying the action with respect to $\varphi^{\ast}$ and $\varphi$ we get
the Klein-Gordon equation
\begin{equation}
\ddot{\varphi}+3H\dot{\varphi}+m^{2}\varphi+2\lambda (\varphi
\varphi^{\ast})\varphi=0~,  \label{kle}
\end{equation}
and its complex conjugate.
The system is fully determined by the independent eqs.(\ref{e00}) 
and (\ref{kle}). In fact, one can easily show that
eq.(\ref{eij}) follows from
eqs.(\ref{e00}) and
(\ref{kle}). 
For the massive scalar field case we use the
following dimensionless variables
\begin{eqnarray}
\begin{array}{l}
t\rightarrow \eta=mt~,\\
\lambda\rightarrow \Lambda=\frac{\lambda}{8\pi Gm^{2}}~,\\
\varphi\rightarrow x_{1}+ix_{2}=
\sqrt{\frac{8\pi G}{3}} \varphi~,\\
\dot{\varphi}\rightarrow y_{1}+iy_{2}=
\sqrt{\frac{8\pi G}{3}} \frac{\dot{\varphi}}{m}~,\\
H\rightarrow z=\frac{H}{m}~,\\
k\rightarrow \tilde{k}=\frac{k}{m^{2}}~.
\end{array}     \label{eq:res}
\end{eqnarray}
This way, we get for eqs.(\ref{e00}) and (\ref{kle}) the following set
\begin{eqnarray}
\frac{\tilde{k}}{a^{2}}+z^{2} & = & y_{1}^{2}+y_{2}^{2}+x_{1}^{2}+x_{2}^{2}+
3\Lambda(x_{1}^{2}+x_{2}^{2})^{2}~, \label{constr} \\
y_{1}^{'} & = &\mbox{} -3zy_{1}-x_{1}-6
\Lambda x_{1}(x_{1}^{2}+x_{2}^{2})~,  \label{y1} \\
x_{1}^{'} & = & y_{1}~, \label{x1} \\
y_{2}^{'} & = & \mbox{}-3zy_{2}-x_{2}-6
\Lambda x_{2}(x_{1}^{2}+x_{2}^{2})~,  \label{y2} \\
x_{2}^{'} & = & y_{2}~, \label{x2}
\end{eqnarray}
where prime means derivative with respect to $\eta$.
The only singular point (defined as the point ($x_1,x_2,y_1,y_2$)
for which the right hand side of eqs.(\ref{constr}) - (\ref{x2}) vanishes),
which we denote by $A$,
lying in a finite region of the phase space (defined by
$\varphi,\dot\varphi$ or equivalently $x_1, x_2, y_1, y_2$)
is the coordinate origin.
Eqs.(\ref{constr}) - (\ref{x2}) are invariant
under the transformations 
\begin{eqnarray}
\begin{array}{llrlr}
a)& x_{1}\rightarrow &\mbox{}-x_{1}&\mbox{ and }~y_{1}\rightarrow
&\mbox{}-y_{1}~, \\
b)& x_{2}\rightarrow &\mbox{}-x_{2}&\mbox{ and }~y_{2}\rightarrow
&\mbox{}-y_{2}~,\\
c)& x_{2}\leftrightarrow &x_{1}&\mbox{ and }~y_{2} \leftrightarrow &y_{1}~.\\
\end{array} \label{trans2}
\end{eqnarray}
For every solution which describes an expanding Universe
(i.e. $z>0$) there is a
corresponding solution describing a collapsing Universe.
This can be seen by performing one of the following transformations 
on the set of eqs.(\ref{constr}) - (\ref{x2})
\begin{eqnarray}
a)~ \left\{\begin{array}{l}
\eta\rightarrow -\eta\\
z\rightarrow -z\\
x_{1}\rightarrow -x_{1}\\
x_{2}\rightarrow -x_{2}~
\end{array} \right.
,~b)~ \left\{\begin{array}{l}
\eta\rightarrow -\eta\\
z\rightarrow -z\\
x_{1}\rightarrow -x_{1}\\
y_{2}\rightarrow -y_{2}~
\end{array} \right.
,~c)~ \left\{\begin{array}{l}
\eta\rightarrow -\eta\\
z\rightarrow -z\\
y_{1}\rightarrow -y_{1}\\
x_{2}\rightarrow -x_{2}~
\end{array}\right.
,~d)~ \left\{\begin{array}{l}
\eta\rightarrow -\eta\\
z\rightarrow -z\\
y_{1}\rightarrow -y_{1}\\
y_{2}\rightarrow -y_{2}~
\end{array}\right. .     \label{trans3}
\end{eqnarray}
 
For the massless scalar field case we use the
dimensionless variables
\begin{eqnarray}
\begin{array}{l}
t\rightarrow \eta_0=\frac{t}{\sqrt{8\pi G}}~,\\
\varphi\rightarrow x_{10}+ix_{20}=
\sqrt{\frac{8\pi G}{3}} \varphi~,\\
\dot{\varphi}\rightarrow y_{10}+iy_{20}=
\frac{8\pi G}{\sqrt{3}} \dot{\varphi}~,\\
H\rightarrow z_0=\sqrt{8\pi G}H~,\\
k\rightarrow \tilde{k}_0=8\pi G~k~,
\end{array}     \label{eq:res0}
\end{eqnarray}
and $\lambda$ remains unchanged.
This way, we get for eqs.(\ref{e00}) and (\ref{kle}) the set
\begin{eqnarray}
\frac{\tilde{k}_0}{a^{2}}+z_0^{2} & = & y_{10}^{2}+y_{20}^{2}+
3\lambda(x_{10}^{2}+x_{20}^{2})^{2}~, \label{constr0} \\
y_{10}^{'} & = &\mbox{} -3z_0y_{10}-6 \lambda x_{10}(x_{10}^{2}+x_{20}^{2})~,  
\label{y10} \\
x_{10}^{'} & = & y_{10}~, \label{x10} \\
y_{20}^{'} & = & \mbox{}-3z_0y_{20}-6 \lambda x_{20}(x_{10}^{2}+x_{20}^{2})~,  
\label{y20} \\
x_{20}^{'} & = & y_{20}~, \label{x20}
\end{eqnarray}
where prime means here derivative with respect to $\eta_0$.
These equations are also invariant under the transformations 
given by eqs.(\ref{trans2}) and (\ref{trans3}), and the coordinate origin is
the only 
singular point lying in the
finite region of the phase space.\\

\begin{tabbing}
\noindent{\bf 3.} \= {\bf MASSIVE SCALAR FIELD IN A CURVATURELESS
FRIEDMAN-LEMAI-}\\
\> {\bf TRE UNIVERSE}
\end{tabbing}
 
As next we study the asymptotic behavior of the
solutions of eqs.(\ref{constr})-(\ref{x2})
nearby the singular points. Due to the increasing complexity of
the analysis involved for a complex scalar field with respect to a 
real one, we restrict ourselves to the case $k=0$.
In section 5 we briefly comment on the extension to $k= \pm 1$.   
With $\tilde k=0$
in eq.(\ref{constr}) we can then rewrite eqs.(\ref{y1})-(\ref{x2})
in spherical coordinates, defined as follows
\begin{eqnarray}
\begin{array}{l}
x_{1}=r\cos{\vartheta_{3}}\cos{\vartheta_{1}}~,\\
y_{1}=r\cos{\vartheta_{3}}\sin{\vartheta_{1}}~,\\
x_{2}=r\sin{\vartheta_{3}}\cos{\vartheta_{2}}~,\\
y_{2}=r\sin{\vartheta_{3}}\sin{\vartheta_{2}}~,
\end{array}     \label{trans}
\end{eqnarray}
with $\vartheta_{1}$, $\vartheta_{2}\in [0,2\pi)$ and
$\vartheta_{3}\in [0,\pi)$. This way we obtain
\begin{eqnarray}
z^{2}&=&r^{2}+3\Lambda r^{4}(\cos^{2}{\vartheta_{3}}\cos^{2}{\vartheta_{1}}+
\sin^{2}{\vartheta_{3}}\cos^{2}{\vartheta_{2}})^{2}
~,\label{constr2}\\
\vartheta_{1}^{'}&=&-3z\sin{\vartheta_{1}}
\cos{\vartheta_{1}}-1
-6\Lambda r^{2}(\cos^{2}{\vartheta_{3}}
\cos^{2}{\vartheta_{1}}+
\sin^{2}{\vartheta_{3}}\cos^{2}{\vartheta_{2}})
\cos^{2}{\vartheta_{1}}
~,\label{eq1}\\
\vartheta_{2}^{'}&=&-3z
\sin{\vartheta_{2}}\cos{\vartheta_{2}}-1
-6\Lambda r^{2}(\cos^{2}{\vartheta_{3}}
\cos^{2}{\vartheta_{1}}+
\sin^{2}{\vartheta_{3}}\cos^{2}{\vartheta_{2}})
\cos^{2}{\vartheta_{2}}
~,\label{eq2}\\
\vartheta_{3}^{'}&=&\sin{\vartheta_{3}}\cos{\vartheta_{3}}
[-3z(\sin^{2}{\vartheta_{2}}-\sin^{2}{\vartheta_{1}}) \nonumber\\
 & &\mbox{}-6\Lambda r^{2}(\cos^{2}{\vartheta_{3}}
\cos^{2}{\vartheta_{1}}+\sin^{2}{\vartheta_{3}}
\cos^{2}{\vartheta_{2}})
(\sin{\vartheta_{2}}\cos{\vartheta_{2}}-
\sin{\vartheta_{1}}\cos{\vartheta_{1}})]
~,\label{eq3}\\
r^{'}&=&\mbox{}-3rz(\cos^{2}{\vartheta_{3}}\sin^{2}{\vartheta_{1}}+
\sin^{2}{\vartheta_{3}}\sin^{2}{\vartheta_{2}}) \nonumber\\
& &\mbox{}-6\Lambda r^{3}
(\cos^{2}{\vartheta_{3}}\cos^{2}{\vartheta_{1}}+
\sin^{2}{\vartheta_{3}}\cos^{2}{\vartheta_{2}}) \times \nonumber\\
& &\mbox{}(\cos{\vartheta_{1}}\sin{\vartheta_{1}}\cos^{2}{\vartheta_{3}}+
\cos{\vartheta_{2}}\sin{\vartheta_{2}}\sin^{2}{\vartheta_{3}})~.\label{eq4}
\end{eqnarray}
Due to the transformations given in eq.(\ref{trans2}) which leave the
equations invariant, we can
restrict the domain of definition for $\vartheta_{1}$, $\vartheta_{2}$
and $\vartheta_{3}$ respectively to $\left[0,\pi\right)$, 
$\left[0,\pi\right)$ and $\left[\frac{\pi}{2},\pi \right)$.
At the singular point $A$ lying at $r=0$
the above equations reduce to
$\vartheta_{1}^{'}=-1$, $\vartheta_{2}^{'}=-1$,
$\vartheta_{3}^{'}=0$ and $r^{'}=0$.
To get the asymptotic behavior of the solution nearby A we apply the
averaging method (for details see  ref. \cite{arn1}). 
Thus, we have to solve the 
differential equation 
\begin{equation}
r^{'}=-\frac{3}{2}~r^{2}~,
\end{equation}
which is obtained by averaging eq.(\ref{eq4}) over the angular variables. 
We get the asymptotic behavior for the solution
nearby $A$
\begin{eqnarray}
\begin{array}{l}
x_{1}=\frac{2}{3\eta}\cos{\vartheta_{30}}\cos({\eta-\eta_{1}})~,\\
y_{1}=\mbox{}-\frac{2}{3\eta}\cos{\vartheta_{30}}\sin({\eta-\eta_{1}})~,\\
x_{2}=\frac{2}{3\eta}\sin{\vartheta_{30}}\cos({\eta-\eta_{2}})~,\\
y_{2}=\mbox{}-\frac{2}{3\eta}\sin{\vartheta_{30}}\sin({\eta-\eta_{2}})~,
\end{array}     \label{res1}
\end{eqnarray}
where $\vartheta_{30}$, $\eta_{1}$ and $\eta_{2}$ are integration
constants. We see that $A$ is an asymptotically stable winding point.
Setting $\vartheta_{30} = 0$ corresponds to consider only a real scalar
field and we recover the solution discussed by Belinsky et al. 
in ref.\cite{bel1,bel2,bel3}.

We study now all singular points lying at
infinity in phase space. First, we consider the case
with no quartic self-interaction term.\\
 
\noindent{\bf 3.1. Properties of the singular points
lying at infinity for} $\bf{\Lambda=0}$\\
 
In order to find the singular points lying at infinity we perform a 
transformation which maps them
on the boundary of a unit 3-sphere,
defined as follows
\begin{eqnarray}
\left\{\begin{array}{l}
r\rightarrow \rho=\frac{r}{1+r}~,\\
d\eta\rightarrow d\tau=\frac{d\eta}{1-\rho}~,
\end{array}\right.     \label{trans4}
\end{eqnarray}
with $\rho \in $ [0,1).
With this transformation eqs.(\ref{eq1}) - (\ref{eq4})
become
\begin{eqnarray}
\frac{d\rho}{d\tau}&=&-3\rho^{2}(1-\rho)
(\cos^{2}{\vartheta_{3}}\sin^{2}{\vartheta_{1}}+
\sin^{2}{\vartheta_{3}}\sin^{2}{\vartheta_{2}})~,
\label{eq5}\\
\frac{d\vartheta_{1}}{d\tau}&=&-3\rho\sin{\vartheta_{1}}
\cos{\vartheta_{1}}-(1-\rho)~,
\label{eq6}\\
\frac{d\vartheta_{2}}{d\tau}&=&-3\rho\sin{\vartheta_{2}}
\cos{\vartheta_{2}}-(1-\rho)~,
\label{eq7}\\
\frac{d\vartheta_{3}}{d\tau}&=&3\rho\sin{\vartheta_{3}}
\cos{\vartheta_{3}}
(\sin^{2}{\vartheta_{1}}-\sin^{2}{\vartheta_{2}})~. \label{eq8}
\end{eqnarray}
Since the system of differential equations is well defined for $\rho=1$,
we can extend the domain of definition for $\rho$ to the boundary.
This way, we have now a compactified phase space.
At infinity (i.e. setting $\rho=1$ in the above eqs.(\ref{eq5})-(\ref{eq8}))
we find two singular curves and two singular
points, which are given by
\begin{eqnarray}
l_{1}&\mbox{: }&\vartheta_{1}=0~,\vartheta_{2}=0~, 
\frac{\pi}{2} \leq \vartheta_{3} < \pi~, \nonumber \\
l_{2}&\mbox{: }&\vartheta_{1}=\frac{\pi}{2}~,\vartheta_{2}=\frac{\pi}{2}~,
\frac{\pi}{2} \leq \vartheta_{3} < \pi~, \nonumber \\
p_{1}&\mbox{:
 }&\vartheta_{1}=\frac{\pi}{2}~,\vartheta_{2}=0~,\vartheta_{3}=\frac{\pi}{2}~,
\nonumber \\
p_{2}&\mbox{:
 }&\vartheta_{1}=0~,\vartheta_{2}=\frac{\pi}{2}~,\vartheta_{3}=\frac{\pi}{2}
~.\nonumber
\end{eqnarray}
To study the asymptotic behavior of the solutions
nearby $l_{1}$ is rather involved,
due to the fact that every point $l_{0}=(1,0,0,\vartheta_{30})
~\in~l_{1}$ is non hyperbolic
(see ref.\cite{arr1} for details). We first perform a variable
shift in eqs.(\ref{eq5})-(\ref{eq8}) defined as follows: 
$\delta\rho=\rho-1$, $\delta\vartheta_{1}=\vartheta_{1}$, 
$\delta\vartheta_{2}=\vartheta_{2}$ and 
$\delta\vartheta_{3}=\vartheta_{3}-\vartheta_{30}$. We then  
expand the differential equations in the new coordinates up to
third order in a sufficiently small neighbourhood  around the singular 
point $l_{0}$. Next, we define the linear coordinate transformation:
$x=\delta\rho,~y=\mbox{}-\frac{1}{3}\delta\rho+\delta\vartheta_{1},
~v=\mbox{}-\frac{1}{3}\delta\rho+\delta\vartheta_{2}\mbox{ and }w=
\delta\vartheta_{3}$, such that we get a set of differential equations of 
the form
\begin{equation}
\left(\vec{y}\right)^{'}=D\vec{y}+\vec{p}~,
\end{equation}
where
$\vec{y}=(x, y, v, w)$.
$D$ is a diagonal matrix and $\vec{p}$ is a vector whose components
are polynomes in $x$, $y$, $v$ and $w$ containing monomes of degree 2 
and 3. 
We can now use the Poincar\'e-Dulac theorem \cite{arn2} to
classify the monomes in $\vec{p}$ into resonant and 
non-resonant ones. For the non-resonant monomes of
degree $n$, there is a polynomial change of coordinates 
of degree $n$, so
that they are transformed into polynomes of at least degree $n+1$.
This is not the case for the resonant monomes.
The polynomial change of coordinates is found by solving the so-called 
homological
equation (see ref.\cite{arn2} for more details). 
Performing the polynomial change of 
coordinates on the differential equations, the non-resonant terms of degree
2 and 3 become of higher order and are thus neglected.
Retaining only terms up to third order we obtain the set of equations
\begin{eqnarray}
\frac{d\tilde{x}}{d\tau}&=&\frac{1}{3}\tilde{x}^{3}~, \label{reso} \\
\frac{d\tilde{y}}{d\tau}&=&\mbox{}-3\tilde{y}-3\tilde{x}\tilde{y}+
\frac{2}{3}\sin^{2}\vartheta_{30}(\tilde{x}^{2}\tilde{y}
-\tilde{x}^{2}\tilde{v})~,\\
\frac{d\tilde{v}}{d\tau}&=&\mbox{}-3\tilde{v}-3\tilde{x}\tilde{v}-
\frac{2}{3}\cos^{2}\vartheta_{30}(\tilde{x}^{2}\tilde{y}
-\tilde{x}^{2}\tilde{v})~,\\
\frac{d\tilde{w}}{d\tau}&=&0 ~,
\end{eqnarray}
where $(x, y, v, w) = (\tilde{x}, \tilde{y}, 
\tilde{v}, \tilde{w})+$ polynomes of degree 2 or higher
in $(\tilde{x},\tilde y,\tilde v,\tilde w)$. 
Since the solution of eq.(\ref{reso}) does
monotonically increase as a function of $\tau$, $l_{0}$ is a saddle 
point.  
The above equations can now be solved analytically and furthermore by 
transforming back to the original variables 
one gets for the asymptotic behavior nearby any point
$l_{0}$ of $l_{1}$
\begin{eqnarray}
\varphi&=& \frac{-M_{p}mt}{\sqrt{3}}~e^{i\vartheta_{30}}~,\label{eql01} \\
\dot{\varphi}&=& \frac{-M_{p}m}{\sqrt{3}}~e^{i\vartheta_{30}}~,
\label{eql02} \\
H&=& \frac{-m^{2}t}{3}\label{eql03}~,
\end{eqnarray}
for $t\rightarrow\mbox{}-\infty$, corresponding to the initial singularity and
where $M_{p}=\frac{1}{\sqrt{8\pi G}}$.
This solution corresponds to an 
outgoing line, the so-called separatrix. 
Using eq.(\ref{eql03}) and the definition of $H$ we have 
\begin{equation}
\frac{a(t_{f})}{a(t_{i})}=
\exp \left( \frac{1}{6} m^2 (t_i^2 - t_f^2) \right) 
%\left|\frac{\dot{\varphi}(t_{i})+C_{1}}{\dot{\varphi}(t_{f})+C_{1}}\right|^{
%\frac{1}{3}},~
\end{equation}
for $t_{f}>t_{i}$.

The effective equation of state near $l_0$ tends to
$\varepsilon=-p$, where $\varepsilon=T_{00}$ and $p=\frac{1}{3}
\left(T_{11}+T_{22}+T_{33}\right)$. The solutions near the separatrix are
characterized by the fact that $\dot{\varphi}\dot{\varphi}^{\ast}
\ll m^{2}\varphi\varphi^{\ast}$ and the phase $\vartheta_{30}$
being constant. 
If for a time $t_f-t_i$ a trajectory $T$ satisfies  
these two 
conditions and, at time $t_f$, $T$ is near to the separatrix, then
it follows with
eq.(\ref{e00}) that $\varphi\sim\frac{\sqrt{3}M_{p}}{m}
~He^{i\vartheta_{30}}$. Using eq.(\ref{kle}) 
it is easy to show that
\begin{equation}
\frac{a(t_{f})}{a(t_{i})}=
\left|\frac{\dot{\varphi}(t_{i})+C_{1}}{\dot{\varphi}(t_{f})+C_{1}}\right|^{
\frac{1}{3}},~
\end{equation}
for $t_{f}>t_{i}$,
with $C_{1}=\frac{M_{p}m}{\sqrt{3}}~e^{i\vartheta_{30}}$.
Every trajectory which lies close enough to the separatrix will
thus meet the criteria for inflation. 

To establish the asymptotic behavior of the solution 
nearby the singular saddle point $p_{1}$,
the same strategy as for the line $l_{1}$ has to be applied. We therefore 
give here only the result, which is 
\begin{eqnarray}
\varphi&=& \frac{-iM_{p}mt}{\sqrt{3}}~,\label{p12} \\
\dot{\varphi}&=& \frac{-iM_{p}m}{\sqrt{3}}~,\label{p13} \\
H&=& \frac{-m^{2}t}{3}\label{p14}~,
\end{eqnarray}
for $t\rightarrow\mbox{}-\infty$.
This solution corresponds also to an outgoing separatrix. 
A more detailed analysis shows that the singular point can 
only be reached if $\vartheta_{3}\equiv \frac{\pi}{2}$. 
Otherwise, starting from points nearby $p_1$ with $\vartheta_3 \neq \pi
/2$ the phase of the scalar field varies strongly. In this region
of the phase space inflation is driven by the imaginary
part of the scalar field.
Also here the phase of $\varphi$ remains constant along the
separatrix and we get the same equation of state as for the previous case.
Applying the transformations defined in
eq.(\ref{trans2}) on eq.(\ref{p12}), we get instead of $p_1$
the singular point 
$\tilde p_1=(1,0,\frac{\pi}{2},0)$, for which the 
corresponding solution is now real.  
Hence, the analysis
made in ref.\cite{bel1,bel2,bel3} applies here as well.

For all points $b=(1,\frac{\pi}{2},\frac{\pi}{2},\vartheta_{3})$ 
in $l_{2}$ we can directly solve the linearized differential equations.
It turns out that $\vartheta_{3}$ remains constant and that on the
plane $\vartheta_{3}=\vartheta_{30}=const$ 
the solution expands away from $b$. 
Using the inverse transformations of eqs.(\ref{trans4}) and (\ref{eq:res}),
we obtain the asymptotic behavior of the scalar field and of the
Hubble parameter
\begin{eqnarray}
\varphi&=& \frac{M_{p}}{\sqrt{3}}\ln\left(\frac{t}{t_{0}}\right)~
e^{i\vartheta_{30}}~,\label{l21} \\
\dot{\varphi}&=& \frac{M_{p}}{\sqrt{3}t}~ e^{i\vartheta_{30}}~,
\label{l22} \\
H&=& \frac{1}{3t}\label{l23}~,
\end{eqnarray}
for $t\rightarrow 0^{+}$ 
corresponding to the initial cosmological singularity
and where $t_{0}$ is an integration constant.
The equation of state nearby these points tends to $\varepsilon=p$
(i.e. stiff matter). From eqs.(\ref{e00}) and (\ref{eij}) we get,
as long as $\dot{\varphi}\dot{\varphi}^{\ast}\gg m^{2}\varphi\varphi^{\ast}$,
that $\dot{H}=-3H^{2}$. Solving
this differential equations and using the definition of $H$, we obtain
as expected
\begin{equation}
\frac{a(t_{f})}{a(t_{i})}=\left(\frac{t_{f}}{t_{i}}\right)^{\frac{1}{3}}.~
\end{equation}
Following the same method used for the points in $l_{2}$, we get for 
the saddle point $p_{2}$
\begin{eqnarray}
\varphi&=& M_{p}\sqrt{3}\left[C_{2}+i\frac{1}{3}\ln(\frac{t}{t_{0}})\right]~,
\label{p21} \\
\dot{\varphi}&=& M_{p}\sqrt{3}\left[\frac{-C_{2}m^{2}t}{2}
+\frac{i}{3t}\right]~,
\label{p22} \\
H&=& \frac{1}{3t}\label{p23}~,
\end{eqnarray}
for $t\rightarrow 0^{+}$, with $C_{2}$ and  $t_{0}$ being integration 
constants. The results of the analysis for $l_{2}$ do also apply here.

This completes the study of the phase portraits for $\Lambda=0$, for which 
we found two singular curves $l_1$, $l_2$ and two singular points
$p_1$, $p_2$. All other singular curves and points, due to
the transformations given in eq.(\ref{trans2}), can be reduced to one of this
cases.
For the curve $l_1$ and the point $p_1$ the solutions
correspond to an outgoing separatrix where inflation occurs. Along
these separatrices the phase of $\varphi$ remains
constant. 
For the curve $l_2$, for which the effective equation of state corresponds
to stiff matter, the phase remains also constant. 
Setting $\vartheta_{30}=0$ in the solutions around $l_1$ and $l_2$, we recover
the results found by Belinsky et al.\cite{bel1,bel2,bel3} for a
real scalar field.
Contrary to the previous ones
the solution around $p_2$, for which we also get an effective equation of
state for stiff matter, can not
be obtained by simply adding a constant phase to the corresponding
asymptotic solution 
for the real scalar field. 
As next we turn to the case where there is a quartic
self-interaction term. \\
 
\noindent{\bf 3.2. Properties of the singular points
lying at infinity for} $\bf{\Lambda\neq 0}$\\
 
We perform on eqs.(\ref{eq1})-(\ref{eq4}) the
transformation defined by eq.(\ref{trans4}). We then obtain 
the set of 
equations
\begin{eqnarray}
\frac{d\rho}{d\tau}&=&-3\rho^{2}f
(\cos^{2}{\vartheta_{3}}\sin^{2}{\vartheta_{1}}+
\sin^{2}{\vartheta_{3}}\sin^{2}{\vartheta_{2}}) \nonumber \\
&
 &\mbox{}-6\Lambda\rho^{3}g(\sin{\vartheta_{1}}\cos{\vartheta_{1}}
\cos^{2}{\vartheta_{3}}+
\sin{\vartheta_{2}}\cos{\vartheta_{2}}\sin^{2}{\vartheta_{3}})~,
\label{eq9}\\
\frac{d\vartheta_{1}}{d\tau}&=&-\frac{1}{1-\rho}\left[
(1-\rho)^{2}+3\rho f\sin{\vartheta_{1}}\cos{\vartheta_{1}}+
6\Lambda\rho^{2}g\cos^{2}{\vartheta_{1}}\right]~,
\label{eq10}\\
\frac{d\vartheta_{2}}{d\tau}&=&-\frac{1}{1-\rho}\left[
(1-\rho)^{2}+3\rho f\sin{\vartheta_{2}}\cos{\vartheta_{2}}+
6\Lambda\rho^{2}g\cos^{2}{\vartheta_{2}}\right]~,
\label{eq11}\\
\frac{d\vartheta_{3}}{d\tau}&=&-\frac{1}{1-\rho}\left[
3\rho f\sin{\vartheta_{3}}\cos{\vartheta_{3}}(\sin^{2}{\vartheta_{2}}-
\sin^{2}{\vartheta_{1}})\right. \nonumber \\
& &\left. \mbox{}+6\Lambda\rho^{2}g\cos{\vartheta_{3}}\sin{\vartheta_{3}}
(\cos{\vartheta_{2}}\sin{\vartheta_{2}}-
\cos{\vartheta_{1}}\sin{\vartheta_{1}})\right],~
\label{eq12}
\end{eqnarray}
where
\begin{equation}
f=\sqrt{(1-\rho)^{2}+3\Lambda\rho^{2}g^{2}}
\end{equation}
and
\begin{equation}
g=\cos^{2}{\vartheta_{3}}\cos^{2}{\vartheta_{1}}+
\sin^{2}{\vartheta_{3}}\cos^{2}{\vartheta_{2}}~. \label{g}
\end{equation}
 
Since the right-hand side of eq.(\ref{eq9}) is well defined and continuous 
for $\rho=1$,
we can find all sets of angles $(\vartheta_{10},\vartheta_{20},
\vartheta_{30})$ for which the condition $\frac{d\rho}{d\tau}=0$ at 
$\rho=1$ is fulfilled. For each solution $(\vartheta_{10},\vartheta_{20},
\vartheta_{30})$ we have to check if
$\lim_{\rho {\rightarrow} 1^-} f_{i}
(\rho,\vartheta_{10},\vartheta_{20},\vartheta_{30})=0$,
where $f_{i}$ stands for the right-hand side of 
eqs.(\ref{eq10}) - (\ref{eq12}). If this is the case, 
the point $(1,\vartheta_{10},\vartheta_{20},\vartheta_{30})$ is
a singular point at infinity. 
Again, we find two singular curves and two singular points
\begin{eqnarray}
L_{1}&\mbox{: }&\vartheta_{1}=\arctan\left(\frac{-2\sqrt{3\Lambda}}{3}
\right)+\pi~,
\vartheta_{2}=\arctan\left(\frac{-2\sqrt{3\Lambda}}{3}\right)+\pi~,
\frac{\pi}{2} \leq \vartheta_{3} < \pi~, \nonumber \\
& &\rho\rightarrow 1~, \label{L1} \\
L_{2}&\mbox{: }&\vartheta_{1}=\frac{\pi}{2}~,\vartheta_{2}=\frac{\pi}{2}~,
\frac{\pi}{2} \leq \vartheta_{3} < \pi~, \rho\rightarrow 1~, \label{L2} \\
P_{1}&\mbox{: }&\vartheta_{1}=\frac{\pi}{2},\vartheta_{2}=
\arctan\left(\frac{-2\sqrt{3\Lambda}}{3}\right)+\pi~,
~\vartheta_{3}=\frac{\pi}{2}~,
\rho\rightarrow 1~, \label{P1} \\
P_{2}&\mbox{: }&\vartheta_{1}=0~,\vartheta_{2}=\frac{\pi}{2}~,
\vartheta_{3}=\frac{\pi}{2}~,
\rho\rightarrow 1~. \label{P2}
\end{eqnarray}
Notice that the limit $\Lambda \rightarrow 0$ is rather
subtle, since in the 
differential equations there are terms involving $\Lambda/(1-\rho)$,
which are not defined when $\rho \rightarrow 1$ and 
$\Lambda \rightarrow 0$ simultaneously.
We expand the differential equations around an arbitrary point
$P_{0}$ of $L_{1}$ 
defined by the following coordinates
\begin{displaymath}
\rho,\vartheta_{1}=\arctan\left(\frac{-2\sqrt{3\Lambda}}{3}\right)+\pi,
\vartheta_{2}=\arctan\left(\frac{-2\sqrt{3\Lambda}}{3}\right)+\pi,
\vartheta_{3}=\vartheta_{30}~,
\end{displaymath}
where $\vartheta_{30}$ is in $[\frac{\pi}{2},~\pi)$.
In order to have finite partial derivatives, when $\rho$ tends
to $1$, the angular variables $\vartheta_1$ and $\vartheta_2$ are kept fixed. 
When linearizing eq.(\ref{eq12}) around $P_0$, we see that also the
angular variable $\vartheta_3$ has to remain constant. Inserting the values 
of the angular variables defined in eq.(\ref{trans}), 
we get for the scalar field
\begin{eqnarray}
\varphi&=&\frac{-3M_{p}m~e^{i\vartheta_{30}}}
{\sqrt{4\lambda M_{p}^{2}+3m^{2}}}~r~,\label{eqL11}\\
\dot{\varphi}&=&\mbox{}-2\sqrt{\frac{\lambda}{3}}~M_{p}\varphi~,
\label{eqL12}
\end{eqnarray}
for $r \rightarrow \infty$, where r is a dimensionless
parameter defined by eq.(\ref{trans}). 
Since the phase of $\varphi$ is constant, 
we can integrate eq.(\ref{eqL12}) and thus get $\varphi$ as a function of $t$
rather than $r$. This way we get
\begin{equation}
\varphi= \varphi_0 e^{i\vartheta_{30}}
\exp\left(-2M_p\sqrt{\frac{\lambda}{3}}~t\right)~, \label{LL1}
\end{equation}
where $t\rightarrow -\infty$ and $\varphi_0$ is a negative integration 
constant. 
Using eq.(\ref{e00}), one gets the
asymptotic behavior for the Hubble parameter. This solution corresponds
to an outgoing separatrix and the equation of state tends to $\varepsilon=-p$.
A trajectory, which lies sufficiently close to the separatrix, can be
parametrized by $\dot{\varphi}\sim\alpha\varphi$, where $\alpha\sim
-2M_p\sqrt{\lambda /3}$. 
As long as $H^{2}\sim \frac{1}{3M_{p}^{2}}\lambda (\varphi\varphi^{\ast})^{2}$,
with eq. (\ref{kle}) we find along the trajectory
\begin{equation}
\frac{a(t_{f})}{a(t_{i})}=
\exp \left( \frac{-\alpha^2}{3\alpha+
2\sqrt{3\lambda}M_{p}}(t_f-t_i)\right)~,
\end{equation}
for $t_{f}>t_{i}$. The factor in the exponential multiplying 
the time difference is just
the Hubble expansion rate, which is positive and tends to infinity as
$\alpha$ reduces to $-2M_P\sqrt{\lambda / 3}$. Thus, any solution which
lies sufficiently close to the separatrix will go through
an inflationary stage.

Using the same strategy as before for the point $P_{1}$, we obtain
\begin{eqnarray}
\varphi= i\varphi_0 
\exp\left(-2M_p\sqrt{\frac{\lambda}{3}}~t\right)~, 
\label{eqP11}
\end{eqnarray}
where $t\rightarrow -\infty$ and $\varphi_0$ is a negative
integration constant. This solution corresponds also to an outgoing
separatrix. The singular point can only be reached if
$\vartheta_{3}\equiv \frac{\pi}{2}$, thus the real part of the
scalar field vanishes. 
The analysis made for $L_{1}$ applies as well, again there is 
an inflationary stage if the solution gets close to the separatrix. 

For the lines $L_{2}$ we expand the set of
differential equations to first order around a point $B$ defined by the
coordinates $(\rho,~\frac{\pi}{2},~\frac{\pi}{2},~\vartheta_{30})$. For
$\rho \rightarrow 1$ the linearized equations reduce to those obtained
for the line $l_{2}$. Therefore, the solutions are given by eqs.(\ref{l21}) -
(\ref{l23}). The inclusion of a quartic term in the lagrangean 
does not effect
the singular line. This is expected, since near $L_{2}$ the
potential term is negligible compared to the kinetic term
$\dot{\varphi}\dot{\varphi}^{\ast}$. The same remark holds for the 
singular point $P_{2}$, so that the asymptotic behavior is given by
eqs.(\ref{p21})-(\ref{p23}). 

We see that the presence of a quartic self-interaction term does not 
change substantially the main features of the phase portrait. We find again two
singular curves and two singular points. 
All other solutions can be reduced to these by the transformations
given in eq.(\ref{trans2}).
The solutions which correspond
to outgoing separatrices show an inflationary stage. The behavior of
the solutions around these separatrices is not much 
affected by the quartic term. Its
main influence is to shift the position of the 
singular points, where inflation occurs. For the solutions for which  
no inflation occurs, that is around $L_2$ and $P_2$, the results are 
the same as for $l_1$ and $p_2$ respectively. We obtain the
solutions of Belinsky el al. in ref.\cite{bel2} for the real scalar field by 
setting
$\vartheta_3=0$ as mentioned in section 3.1.

In Fig.1 and 2 we plot the numerical
solutions of the differential equations (\ref{y1}) - (\ref{x2})
for different values of
$\Lambda$. For all solutions we take the same
initial conditions as given by the asymptotic behavior near 
the singular point $p_{2}$.
The plots are valid for every value of $m \neq 0$, since 
eqs.(\ref{y1}) - (\ref{x2}) do not depend on it.
The numerical solutions have only a physical
meaning in the region where $T_{00}<M_{p}^{4}$. 
Hence, depending on the value of $m$, it might be that only a part of
the Figure is of physical relevance.

As next, we analyse the asymptotic behavior of the solutions of 
eqs.(\ref{y10})-(\ref{x20}) for a massless scalar field with a quartic
self-interaction nearby the singular points.\\

\begin{tabbing}
\noindent{\bf 4.} \= {\bf MASSLESS SCALAR FIELD IN A CURVATURELESS
FRIEDMAN-LE-}\\
\> {\bf MAITRE UNIVERSE}
\end{tabbing}

Again the coordinate 
origin is the only singular point lying in the finite region of the
phase space. This point is asymptotically stable, since we have the
following Liapunov function (see ref.\cite{guck} for details)
\begin{displaymath}
l(x_{10},x_{20},y_{10},y_{20})=y_{10}^2+y_{20}^2+3\lambda
(x_{10}^2+x_{20}^2)^2~,
\end{displaymath}
for which the derivative with respect to $\eta_0$ is strictly negative,
except at the origin. Applying the transformation defined in
eq.(\ref{trans}) on
eqs.(\ref{constr0})-(\ref{x20}), we get
\newpage
\begin{eqnarray}
z_0^{2}&=&r^{2}\left\{ 1+\left[ 3\lambda r^{2}
(\cos^{2}{\vartheta_{3}}\cos^{2}{\vartheta_{1}}+
\sin^{2}{\vartheta_{3}}\cos^{2}{\vartheta_{2}})
-1\right]\right. \times \nonumber\\
 & & \left.(\cos^{2}{\vartheta_{3}}\cos^{2}{\vartheta_{1}}+
\sin^{2}{\vartheta_{3}}\cos^{2}{\vartheta_{2}})\right\}
~,\label{constr20}\\
\vartheta_{1}^{'}&=&-3z_0\sin{\vartheta_{1}}
\cos{\vartheta_{1}}
-6\lambda r^{2}(\cos^{2}{\vartheta_{3}}
\cos^{2}{\vartheta_{1}}+
\sin^{2}{\vartheta_{3}}\cos^{2}{\vartheta_{2}})
\cos^{2}{\vartheta_{1}}-\sin^{2}{\vartheta_{1}}
~,\label{eq1110}\\
\vartheta_{2}^{'}&=&-3z_0
\sin{\vartheta_{2}}\cos{\vartheta_{2}}
-6\lambda r^{2}(\cos^{2}{\vartheta_{3}}
\cos^{2}{\vartheta_{1}}+
\sin^{2}{\vartheta_{3}}\cos^{2}{\vartheta_{2}})
\cos^{2}{\vartheta_{2}}-\sin^{2}{\vartheta_{2}}
~,\label{eq20}\\
\vartheta_{3}^{'}&=&\sin{\vartheta_{3}}\cos{\vartheta_{3}}
[-3z_0(\sin^{2}{\vartheta_{2}}-\sin^{2}{\vartheta_{1}}) \nonumber\\
 & &\mbox{}-6\lambda r^{2}(\cos^{2}{\vartheta_{3}}
\cos^{2}{\vartheta_{1}}+\sin^{2}{\vartheta_{3}}
\cos^{2}{\vartheta_{2}})
(\sin{\vartheta_{2}}\cos{\vartheta_{2}}-
\sin{\vartheta_{1}}\cos{\vartheta_{1}}) \nonumber\\
 & &\mbox{}+ (\sin{\vartheta_{2}}\cos{\vartheta_{2}}-
\sin{\vartheta_{1}}\cos{\vartheta_{1}})] 
~,\label{eq30}\\
r^{'}&=&\mbox{}-3rz_0(\cos^{2}{\vartheta_{3}}\sin^{2}{\vartheta_{1}}+
\sin^{2}{\vartheta_{3}}\sin^{2}{\vartheta_{2}}) \nonumber\\
& &\mbox{}+r(\cos{\vartheta_{1}}\sin{\vartheta_{1}}\cos^{2}{\vartheta_{3}}+
\cos{\vartheta_{2}}\sin{\vartheta_{2}}\sin^{2}{\vartheta_{3}})
\nonumber\\
& &\mbox{}-6\lambda r^{3}
(\cos^{2}{\vartheta_{3}}\cos^{2}{\vartheta_{1}}+
\sin^{2}{\vartheta_{3}}\cos^{2}{\vartheta_{2}}) \times \nonumber\\
& &(\cos{\vartheta_{1}}\sin{\vartheta_{1}}\cos^{2}{\vartheta_{3}}+
\cos{\vartheta_{2}}\sin{\vartheta_{2}}\sin^{2}{\vartheta_{3}})
~.\label{eq40}
\end{eqnarray}
In a sufficiently small neighbourhood around the origin, which we denote by W,
we consider the projection of the solutions of eqs.(\ref{eq1110})-(\ref{eq40})
on the $(x_{10},~y_{10})$-plane.
The angular variable on this plane is $\vartheta_1$ and
its behavior is given by the
solution of the equation $\vartheta_1^{'}=-\sin^2{\vartheta_1}$, 
which is just eq.(\ref{eq1110}) with $r=0$. 
Therefore, knowing that the coordinate origin is asymptotically stable and
that $\vartheta_1^{'}$ is almost everywhere strictly negative,
it follows that the solutions are winding towards
the point $A_0=(x_{10}=0,~y_{10}=0)$.
One gets 
the same behavior
when the solutions of eqs.(\ref{eq1110})-(\ref{eq40}) in W are projected
on the
$(x_{20},~y_{20})$-plane. 

We now turn to the singular points lying at infinity. One has to
apply on eqs.(\ref{eq1110})-(\ref{eq40}) the transformations given in 
eq.(\ref{trans4}), 
where $\eta$ has to be replaced by $\eta_0$. We obtain
the following equations
\begin{eqnarray}
\frac{d\rho}{d\tau}&=&-3\rho^{2}f
(\cos^{2}{\vartheta_{3}}\sin^{2}{\vartheta_{1}}+
\sin^{2}{\vartheta_{3}}\sin^{2}{\vartheta_{2}}) \nonumber \\
& &\mbox{}+\left[-6\lambda\rho^{3}g+\rho(1-\rho)^2\right]\times \nonumber \\
& & \left[\sin{\vartheta_{1}}\cos{\vartheta_{1}}
\cos^{2}{\vartheta_{3}}+
\sin{\vartheta_{2}}\cos{\vartheta_{2}}\sin^{2}{\vartheta_{3}}\right]~,
\label{eq90}\\
\frac{d\vartheta_{1}}{d\tau}&=&\mbox{}-\frac{1}{1-\rho}
\left[3\rho f\sin{\vartheta_{1}}\cos{\vartheta_{1}}+
6\lambda\rho^{2}~g\cos^{2}{\vartheta_{1}}
+(1-\rho)^2\sin^2{\vartheta_1}\right]~,
\label{eq100}\\
\frac{d\vartheta_{2}}{d\tau}&=&\mbox{}-\frac{1}{1-\rho}
\left[3\rho f\sin{\vartheta_{2}}\cos{\vartheta_{2}}+
6\lambda\rho^{2}~g\cos^{2}{\vartheta_{2}}
+(1-\rho)^2\sin^2{\vartheta_2}\right]~,
\label{eq110}\\
\frac{d\vartheta_{3}}{d\tau}&=&\mbox{}-\frac{
\cos{\vartheta_3}\sin{\vartheta_3}}{1-\rho}\left\{
3\rho f(\sin^{2}{\vartheta_{2}}-\sin^{2}{\vartheta_{1}})\right. \nonumber \\
& &\mbox{}+\left. \left[6\lambda\rho^{2}~g-(1-\rho)^2\right] 
(\cos{\vartheta_{2}}\sin{\vartheta_{2}}-
\cos{\vartheta_{1}}\sin{\vartheta_{1}})\right\}~,
\label{eq120}
\end{eqnarray}
where from now on 
\begin{equation}
f=\sqrt{(1-\rho)^2(1-g)+3\lambda\rho^2g^2}
\end{equation}
and g is still given by eq.(\ref{g}). 
The singular points lying at infinity of the phase space are found using 
the same method as for $\Lambda \neq 0$. We get again two 
singular curves and two singular points, which we denote by
$l_{10}$, $l_{20}$, $p_{10}$ and $p_{20}$ and their coordinates 
are given 
by eqs.(\ref{L1})-(\ref{P2}), but now with $\Lambda$ replaced by 
$\lambda$. To obtain the asymptotic behavior around these singular points
we apply the same method used in section 3.2.

The asymptotic behavior around a singular point lying in
$l_{10}$ is
given by eqs.(\ref{eqL11})-(\ref{eqL12}), but where $m$ is now
replaced by $M_p$. 
The angular variables must be kept fixed in order to have finite partial 
derivatives, when $\rho \rightarrow 1$. The analysis made for $L_1$ is 
also  valid 
here and $l_{10}$ has 
inflationary stages. Setting $\vartheta_3=0$
we obtain automatically the asymptotic behavior for a 
massless real scalar field with a quartic self-interaction for which
we recover
the inflationary stage. A fact this
which was established heuristically by Linde in ref.\cite{lin}.

One gets the behavior of the solutions around the singular point 
$p_{10}$ from the one near $P_1$ in the same way as discussed
above for $l_{10}$ from $L_1$.

The asymptotic behavior of the solutions around all points
$c=(1,\frac{\pi}{2},\frac{\pi}{2},\vartheta_{30})$ of $l_{20}$
or around the point
$p_{20}$ is found directly by solving the corresponding
linearized
differential equations. For the singular points on $l_{20}$
it turns out that 
the solutions are given by eqs.(\ref{l21})-(\ref{l23}), whereas
for $p_{20}$
we get the solutions from eqs.(\ref{p21})-(\ref{p23}) inserting
$m=0$.
The form of the solutions around the line $l_{20}$ and $p_{20}$
was expected to be similar to that obtained for $l_2$ and $p_2$, because
the potential term is negligible with respect to the 
kinetic energy term $\dot{\varphi}\dot{\varphi}^*$.
\\ 

\noindent{\bf 5. CONCLUDING REMARKS}\\

The extension of the above analysis to the singular points for 
$k=\pm 1$, although in principle straightforward, is much more involved. 
One could, for instance,
using eq.(\ref{e00}) 
eliminate the curvature term in eq.(\ref{eij}) and consider this
modified equation together with eq.(\ref{kle}). This gives then a set of
five non-linear first order differential equations. As a consequence, 
when performing the transformation to spherical coordinates needed in
order to compactify the phase space, one gets one additional angular
variable. It turns out that around some singular points the expansion
of the differential equations must be done at least up to fourth order.
The only singular point lying in the finite region of the phase space is
at the coordinate origin. We conjecture that the asymptotic behavior
near the singular points lying at infinity with $k \neq 0$ will not
fulfill the criteria for inflation. 
This does not imply
that inflation can not occur in an open or a 
closed universe, but that every trajectory 
must come close enough to one of the separatrix found for $k = 0$
in order to go through  an inflationary stage.
This fact has been shown for
the real scalar field case (see ref.\cite{bel1,bel2,bel3}). 

In this paper we have extended to complex scalar fields the analysis
of the initial conditions 
in an homogeneous and isotropic Friedman-Lema\^{\i}tre
Universe. The main features found for real scalar fields hold also for 
complex scalar fields, in particular the existence of inflationary stages.
The fact that along the
separatrices  the phase of $\varphi$ remains constant is important and
shows that inflation is essentially driven by one component
of the complex scalar field.
Therefore, the results on inflation valid for a real scalar 
field (see for instance ref.\cite{oli} and references therein) 
apply also on the component of the complex fields which drives
inflation. 
The behavior around the singular points $p_{2}$, $p_{20}$
and $P_{2}$ is more involved 
and can
not be obtained by just adding a phase to the corresponding
solutions for the real scalar field. We also
notice that for a massive scalar field the presence of a quartic 
self-interaction term does not change substantially
the main features of the phase 
portrait.\\

\noindent{\bf Acknowledgements}\\

We thank N. Straumann for very useful and clarifying discussions.\\

\begin{figcap}
\item Plot of the dimensionless variable 
      $y_{1}$ as function of $x_{1}$ for different values of $\Lambda$.
      It corresponds  to the phase portrait of the real part of
      the scalar field.
      All curves start with the same initial condition
      given by the asymptotic behavior near $p_{2}$ (see eq. (\ref{p21})).
      The real part of the scalar fields 
      depends on the value of $C_{2}$, for which
      we take the value $C_{2}=1$.
   
\item Plot of the phase portrait of the imaginary part 
      of the scalar field in dimensionless variables 
      for the same cases as in Fig.1
      The asymptotic behavior of the imaginary part near
      $p_{2}$ depends on the value of $t_{0}$, for which we choose
      $t_{0}=100$.

\end{figcap}

\begin{thebibliography}{99}
\bibitem{kn:Smoot} G.F. Smoot et al. ApJ. {\bf 396}, L1 (1992).
\bibitem{kn:Jetzer} P. Jetzer, Phys. Rept. {\bf 220}, 163 (1992).
\bibitem{kn:Liddle} A. Liddle and M. Madsen, Int. J. of Mod. Phys. D{\bf 1},
101 (1992). 
\bibitem{bel1} V. A. Belinsky, L. P. Grishchuk, I. M. Khalatnikov and
Ya. B. Zeldovich, Phys. Lett. {\bf B155}, 232 (1985).
\bibitem{bel2} V. A. Belinsky, L. P. Grishchuk, 
Ya. B. Zeldovich and I. M. Khalatnikov, Sov. Phys. JETP  {\bf 62} 195 (1985).
\bibitem{bel3} V. A. Belinsky and I. M. Khalatnikov, 
Sov. Phys. JETP  {\bf 66} 441 (1987).
\bibitem{kn:Piran} T. Piran and R.M. Williams, Phys. Lett. {\bf B163},
331 (1985). 
\bibitem{kn:Deruelle} N. Deruelle, C. Gundlach and D. Langlois,
Phys. Rev. D{\bf 45}, 3301 (1992); N. Deruelle, C. Gundlach and D. Polarski,
Class. Quantum Grav. {\bf 9}, 1511 (1992). 
\bibitem{kn:Mukhanov} V. Mukhanov, H. Feldman and R. Brandenberger,
Phys. Rept. {\bf 215}, 203 (1992).
\bibitem{arn1} V. I. Arnold, {\it Geometrical Methods in Theory of
Differential Equations} (Springer-Verlag, New-York, 1983) p.142
\bibitem{arr1} D. K. Arrowsmith and C. M. Place, {\it An Introduction 
to Dynamical System} (Cambridge University Press, Cambridge, 1990) p.79
\bibitem{arn2} V. I. Arnold, {\it Geometrical Methods in Theory of
Differential Equations} (Springer-Verlag, New-York, 1983) p.191
\bibitem{guck} J. Guckenheimer and P. Holmes, {\it Nonlinear Oscillations,
Dynamical Systems and Bifurcations of Vector Fields} 
(Springer-Verlag, New-York, 1983) p.4
\bibitem{lin} A. D. Linde, Phys. Lett. {\bf B129}, 177 (1983).
\bibitem{oli} K. A. Olive, Phys. Rep. {\bf 190}, 307 (1990).
\end{thebibliography}
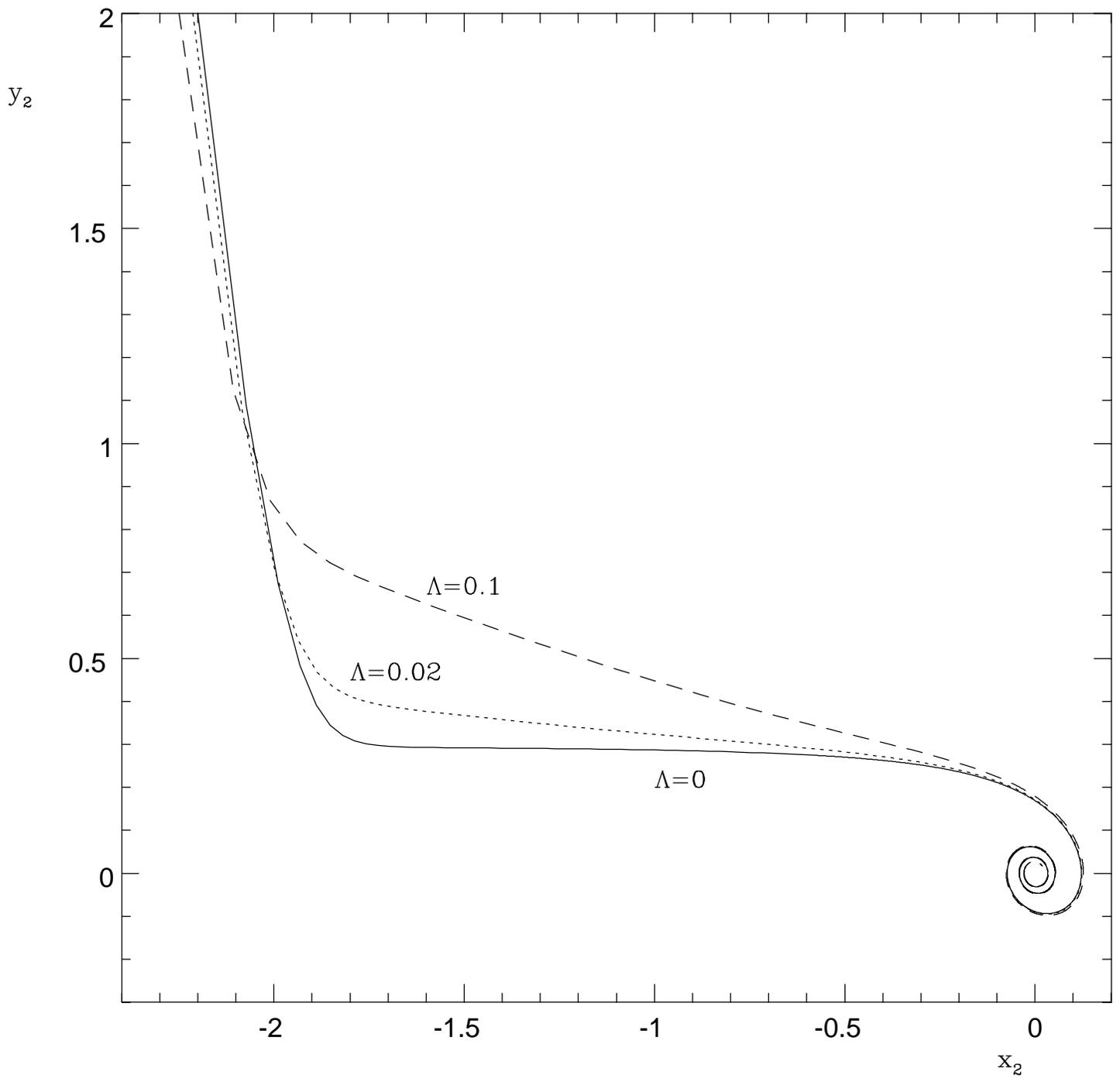
\end{document}